\documentclass[twocolumn]{aastex63}
\usepackage{graphicx}
\usepackage{txfonts}
\usepackage{setspace} 
\usepackage[utf8]{inputenc}
\usepackage[inline, shortlabels]{enumitem}

\def\be{\begin{equation}}
\def\bea{\begin{eqnarray}}
\def\eea{\end{eqnarray}}
\def\ee{\end{equation}}

\newcommand{\beq}{\begin{equation}}
\newcommand{\eeq}{\end{equation}}
\newcommand{\ba}{\begin{array}}
\newcommand{\ea}{\end{array}}

\submitjournal{ApJL}

\shorttitle{Radiation driven magnetic fields}
\shortauthors{Vyas and Pe'er}
\graphicspath{{./}{figures/}}

\begin{document}
\setstretch{1.0}
\title[Radiation triggered magnetic fields around black holes]{Generation of magnetic fields around black hole accretion discs due to non conservative radiation fields}

\correspondingauthor{Mukesh Kumar Vyas} 
\email{mukeshkvys@gmail.com}


\author{Mukesh Kumar Vyas}
\affiliation{Bar Ilan University, \\ Ramat Gan, Israel,
 5290002}
\author{Asaf Pe'er}
\affiliation{Bar Ilan University, \\ Ramat Gan, Israel,
 5290002}

\begin{abstract}
We investigate the generation of magnetic fields above black hole accretion discs due to the non-zero curl of the disc radiation field. By self consistently computing the components of the radiation flux and their curl, we show that the rotational nature of the radiation field induces charge separation, leading to magnetic field generation in the plasma above the disc. Solving the magnetohydrodynamic equations, we derive the time evolution of these fields and demonstrate that they grow over astrophysically relevant timescales. 
For a standard Keplerian accretion disc, the produced magnetic fields remain weak, on the order of a few Gauss, consistent with previous predictions. However, when a luminous corona is present in the inner disc region ($r_d < 3-10 r_g$), the generated fields reach dynamically significant strengths of up to $10^5$ Gauss where the magnetic energy density approaching few percentage of equipartition with the gas pressure. These fields develop within realistic growth timescales (such as viscous timescale) and can be dynamically significant in governing disc and jet evolution. 
Our findings suggest that radiation-driven magnetic fields play a crucial role in accretion flow magnetization, influencing both disc dynamics and observational signatures. The predicted field strengths could affect the thermal emission, synchrotron radiation, and polarization properties of black hole accretion systems, with implications for X-ray binaries, AGN, and jet formation. Future numerical simulations and high-resolution polarimetric observations, such as those from IXPE, eXTP, and EHT, may provide observational confirmation of our findings. 

\end{abstract}

\section{Introduction}
Magnetic fields are ubiquitously present in the astrophysical arena with their generation and growth being an intriguing problem \citep{1988ASSL..133.....R, 2002RvMP...74..775W, 2002PhT....55l..40K, 2005PPCF...47A.205S, 2015ApJ...808...65F, kato2020fundamentals}.  
In the context of X-ray binaries, structured poloidal and toroidal magnetic fields are key components of modern-age magnetohydrodynamic simulations \citep{2021ApJ...914...55W, 2022MNRAS.511.3795N, 2023ApJS..264...32B}; for reviews, see, e.g., \citet{2020ARA&A..58..407D, 2021NewAR..9201610K}. 

There are numerous mechanisms to produce and amplify magnetic fields in astrophysical plasmas. Notable of which include: (i) dynamo process in celestial bodies, that takes place through the motion of electrically conducting fluids, converts kinetic energy into magnetic energy via induction \citep{1978mfge.book.....M}; (ii) plasma instabilities \citep{2005PPCF...47A.205S} such as Weibel instability \citep{1959PhRvL...2...83W} and magneto-rotational instability \cite[MRI,][]{1991ApJ...376..214B,1995ApJ...440..742H} that strongly amplify seed magnetic fields; (iii) the Biermann battery process \citep{1997ApJ...480..481K, 2011ApJ...741...93D} where a misalignment between temperature and pressure gradients leads to non zero currents in the plasma and thus generates magnetic fields \citep{1950ZNatA...5...65B};  (iv) inhomogeneous reionization through photoionization in the early universe \citep{2005A&A...443..367L}; and (v) non-conservative nature of external radiation forces, namely  $\vec \nabla \times \vec F_{\rm rad} \neq 0$ \citep[in the general case; e.g., ][]{2010ApJ...716.1566A, 2014ApJ...782..108S}.  

This last mechanism can be manifested in various ways. For example, the Poynting-Robertson effect, where a central illuminating object causes radiation drag on accretion-disc  electrons, thereby creating charge separation and inducing toroidal currents \citep{1977A&A....59..111B}. This idea was generalized to include finite disc conductivity and illumination from the inner parts of the disc, which is referred to as the ``cosmic battery" \citep{1998ApJ...508..859C, 2002ApJ...580..380B, 2006ApJ...652.1451C, 2008ApJ...674..388C, 2015ApJ...805..105C}. It was shown that in addition to generating magnetic fields in accretion systems, this mechanism can be significant in other scenarios as well, such as magnetic field production in the early universe \citep{2003PhRvD..67d3505L, 2010ApJ...716.1566A}.

The foundational work by \citet{1977A&A....59..111B} proposed that the magnetic fields generated in accretion discs via the Poynting-Robertson effect are weak and do not exceed $\lesssim 100 $ Gauss over the viscous (accretion) time scale.
In order to achieve equipartition between the magnetic and gas energy densities, unrealistically long timescales, significantly longer than the typical accretion time of matter into the central object (presumably a black hole) are required.



This limitation motivated further investigations. One mechanism is based on the cosmic battery effect, first proposed by \citet{1998ApJ...508..859C} and later developed further in \citet{2015ApJ...805..105C}. The cosmic battery mechanism accounts for several additional effects, such as the finite plasma conductivity, motion of the radiating source, and magnetic field amplification by induction.
\citet{2014ApJ...794...27K} examined how the cosmic battery operates in different disc geometries. They demonstrated that the required time to reach significant magnetic fields is hours to days for stellar-mass black holes with sub-Eddington luminosities and $10^5 - 10^6$ years for active galactic nuclei (AGNs). \citet{2015ApJ...805..105C} further refined these ideas by numerically modeling the time-dependent evolution of the cosmic battery effect, confirming that the mechanism can operate efficiently over long timescales but remains dependent on accretion-driven advection to sustain significant fields. However, even with these improvements, the field strength remains several orders of magnitude below the equipartition level.
 

The above mentioned attempts with Poynting Robertson effect and cosmic battery assumed the existence of a luminous accretion disc that extends beyond the innermost stable circular orbit (ISCO). However, these calculations neglect the possibility of an inner corona that may exist at scales comparable to (maybe below) the ISCO. 
It was theoretically demonstrated that such a corona may exist ranging from two to a few tens of Schwarzschild radii \citep[][and references therein]{Miniutti2004}. The corona may approach luminosity comparable to Eddington luminosities, such as shown for GRS 1915+105 by \cite{2012A&A...537A..18M}  \citep[see further][]{2011NewAR..55..166F, 2019BAAS...51c.126K,  lasheras2021extremes, 2023MNRAS.519.2375C}.
In recent years, growing observational evidence has supported the existence of compact coronae, particularly through the interpretation of X-ray spectral lags \citep{2025MNRAS.536.3284U}. Notable examples include observational studies of XTE J1650–500 \citep[][]{Reis2013}, Cygnus X-1 \citep[][]{2022Sci...378..650K}, and GRO J1655–40 \citep[][]{2023MNRAS.525..221R}.

In this paper, we explore the radiation-triggered magnetic fields around black hole accretion discs generated by the non-conservative radiation force field. We work out a detailed field structure and its time evolution as a result of the luminous corona and the disc. As the rotating inner corona possesses high angular momentum close to the black hole and leads to stronger curl of the radiation force applied on the matter at the disc surface, we show that this picture results in a large-scale magnetic field around accretion discs. 
Considering stellar-mass black hole (BH) systems the generated magnetic fields reach few percentage of equipartition, and become dynamically important within the accretion time scale. 

We therefore conclude that within the physically realizable viscous timescale available for the field growth, the produced magnetic field is several orders of magnitude higher than those predicted from the Poynting-Robertson effect around accretion discs \citep{1977A&A....59..111B, 2002ApJ...580..380B}, and is similar in magnitude to the magnetic field that develops due to  magneto-rotational instability (MRI) in the accretion flow. Therefore, we show that the radiation field acts as an additional source of substantial magnetic field alongside MRI. 


 This paper is organized as follows.
In section \ref{sec_ana} below, we describe the mathematical model of magnetic field generation. In section \ref{sec_disc}, we describe the method of calculation of radiative moments and the respective components of the curl of the radiation force. We present results in section \ref{sec_results} and conclude the work in section \ref{sec_conclusions} by highlighting the importance of this work.
\section{Generation of magnetic field in plasma in the presence of an external radiation field}
\label{sec_ana}

We provide in this section a general expression for the rate of magnetic field generation in a plasma element that lies below or above the accretion disc, due to the illumination of radiation originating from the corona and from the disc. This plasma may correspond to the upper or lower edges of the disc itself, it may originate from disc wind, or be composed of clouds that may exist in the vicinity of the accreting system. The expressions in this section are general, while specific spatial setup will be considered in the following sections.

Consider a plasma that is illuminated by an external radiation field. The equation of motion for a force-free electron is
\be 
-e\textbf{E}+\frac{e}{\sigma_c}\textbf{j}+\textbf{f}_{\rm rad}-\frac{\nabla p_e}{n_e}-\frac{e}{c}\textbf{v}\times \textbf{B}+\frac{1}{cn_e}\textbf{j}\times \textbf{B} 
 + \textbf{f}_{\rm g}= 0.
\label{eq_eom1}
\ee 
Here, $e$ is the charge of the electron and $c$ is the light speed, $\bf E$ and $\bf B$ are electric and magnetic fields, respectively. The gas pressure is denoted by $p_e$, $\sigma_c$ is the conductivity, $\textbf{j}[=e(n_p {\bf v}_p-n_e {\bf {v}}_e)]$ is the electric current density with $n_p$ and $n_e$ being the number densities of electrons and protons and ${\bf v}_e, {\bf v}_p$ are their respective speeds. The gravitational force is $\textbf{f}_{\rm g}$ and the radiation force due to an existing external radiation field is $\textbf{f}_{\rm rad}$, and the fluid velocity, ${\bf v}$ is given by
\begin{equation}
  {\bf  v} = \frac{m_p n_p {\bf v}_p + m_e n_e {\bf v}_e}{m_p n_p + m_e n_e} = \frac{m_p {\bf v}_p + m_e {\bf v}_e}{m_p + m_e} ~~~({\rm for~} n_p \approx n_e = n).
\end{equation}

For charge density $\rho[=e(n_p-n_e)]$, the continuity equation reads
\be 
\frac{\partial \rho}{\partial t}+\nabla \cdot \textbf{j} = 0,
\ee 
and the set of Maxwell's equations are 
\be 
\nabla \cdot \textbf{E} = 4\pi \rho, ~~\nabla \cdot \textbf{B} = 0,
\ee 
and 
\be 
\nabla \times \textbf{E} = -\frac{1}{c}\frac{\partial \textbf{B}}{\partial t},~~ \nabla \times \textbf{B} = \frac{4\pi}{c}\textbf{j}+\frac{1}{c}\frac{\partial \textbf{E}}{\partial t}.
\ee 
To obtain the magnetic field evolution, we operate ``$(-4\pi \sigma_c/ec)\nabla \times$" on both sides of equation \ref{eq_eom1} and obtain
\beq
\begin{array}{l} 
-\frac{4\pi}{c}\nabla\times\textbf{j}+\frac{4\pi\sigma_c}{c}\nabla\times\textbf{E}+\frac{4\pi\sigma_c}{c^2}\nabla\times(\textbf{v}\times \textbf{B}) \nonumber \\
-\frac{4\pi \sigma_c}{c^2}\nabla\times\left(\frac{\textbf{j}\times \textbf{B}}{e n_e}\right) 
-\frac{4\pi \sigma_c}{cen_e^2}\nabla n_e \times \nabla p_e-\frac{4\pi \sigma_c}{ce}\nabla\times \textbf{f}_{\rm rad} = 0.
\label{eq_field_eq}
\end{array}
\eeq
Here the gravitational force term vanishes due to its conservative nature.
Using the second and fourth of Maxwell's equations, it reduces to
\beq
\ba{l} 
\frac{1}{c^2}\frac{\partial^2\textbf{B}}{\partial t^2} - \nabla^2 \textbf{B}+\frac{4\pi \sigma_c}{c^2}\frac{\partial \textbf{B}}{\partial t} = \frac{4\pi \sigma_c}{c^2}\nabla\times(\textbf{v}\times \textbf{B})
\nonumber \\ 
-\frac{4\pi \sigma_c}{c^2}\nabla\times\left(\frac{\textbf{j}\times \textbf{B}}{e n_e}\right) 
-\frac{4\pi \sigma_c}{cen_e^2}\nabla n_e \times \nabla p_e-\frac{4\pi \sigma_c}{ce}\nabla\times \textbf{f}_{\rm rad}. 
\label{eq_mag_general}
\ea  
\eeq
Here $4\pi \sigma_c \equiv t'^{-1}$ denotes the inverse timescale over which electrons respond to an external force and settle into their equilibrium velocity. Given the high conductivity of the plasma, this time scale is of the order $10^{-12}$~s \citep{2010ApJ...716.1566A}. 
Considering times $t \gg t'$, the first two terms in the left hand side of Equation \ref{eq_mag_general} are much smaller than the third term, and can therefore be neglected. 
Furthermore, the Hall term (the second term on the right hand side, that contains $\textbf{j}\times \textbf{B}$) can initially be ignored, being a higher-order term in ${\bf B}$\footnote{This assumption fails at later times, when the magnitude of the magnetic field becomes significant. See further discussion below.}. One is left with 
\be 
\frac{\partial \textbf{B}}{\partial t}  = \nabla\times(\textbf{v}\times \textbf{B})-\frac{c}{en_e^2}\nabla n_e \times \nabla p_e-\frac{c}{e}\nabla\times \textbf{f}_{\rm rad}.
\ee 

This is the conventional induction equation with two additional source terms for the magnetic field. 
The first term on the right-hand side is the induction term and is zero for a plasma in hydrostatic equilibrium or negligible for small bulk speeds. The second and third terms on the right-hand side are the only source terms for generating  magnetic fields. The term proportional to $\nabla n_e \times \nabla p_e$ is the Biermann battery term which states that the misalignment of gradients in density and pressure leads to the generation of a magnetic field. 
The last term states that the presence of a non-conservative radiation field would lead to the generation of a magnetic field. Ignoring the Biermann battery process (i.e., assuming that there are no factors that cause misalignment between density and pressure gradient), the source of the magnetic field in a slowly moving plasma ($v \rightarrow 0 $) is written as {\citep{2010ApJ...716.1566A, 2014ApJ...782..108S}, 
\be 
\frac{d \textbf{B}}{d t}  =-\frac{c}{e}\nabla\times \textbf{f}_{\rm rad}.
\ee 
This equation assumes that the illuminated matter is essentially at rest. 

In the context of flows around accretion discs, as we show below, the dominant term in generating strong magnetic fields is radiation from the rapidly rotating inner corona. Assuming a Keplerian motion, the outer parts of the disc rotate at a much slower speed than the inner corona, as $v_K \propto r^{-1/2}$. As for clouds above and below the disc, they can either originate from disc wind, in which case their initial velocities will be similar to the disc outer layers, or have a separate origin, such as companion stellar wind (in the case of X-ray binaries), in which case their velocities will be even lower. We therefore assume in the calculations presented here that the illuminated plasma is motionless. We will relax this assumption in a future work [Vyas \& Pe'er, in prep.]. 

The rotational property or non zero curl of the radiation field induces a net charge flow in the plasma that leads to magnetic field generation. 
The first-order radiation force is \citep{1977A&A....59..111B, 2005MNRAS.356..145C},
\be 
{f^i_{\rm  rad}} = \frac{\sigma_T}{c}F^i.
\label{eq_rad_force}
\ee 
Here $F^i$ are the $i$-th components of the radiative flux, and $\sigma_T$ is the Thomson cross-section.  Hence the total radiative force for non-relativistic plasma is characterized by radiative flux alone. 
Equation \(\ref{eq_rad_force}\) describes the radiation force per electron in 
the plasma above the accretion disc, given a radiation flux \textbf{\textit{F}} from the disc.
The interaction is considered to be dominated by Thomson scattering cross section \citep{1977A&A....59..111B, 1981ApJ...243L.147O, 2008bhad.book.....K}. 

Solving the equation of magnetic field evolution for constant radiation flux over time, one obtains the time evolution of the magnetic field components in cylindrical coordinate system, $\{r, \phi, z\}$,
\be 
B_r(t) = \frac{-\sigma_T t}{e} \left(\frac{1}{r}\frac{\partial {F_z}}{\partial \phi}-\frac{\partial {F_\phi}}{\partial z}\right),
\label{eq_br}
\ee 
\be 
B_\phi(t) = \frac{-\sigma_T t}{e} \left(\frac{\partial {F_r}}{\partial z}-\frac{\partial {F_ z}}{\partial r}\right),
\label{eq_bfi}
\ee 
and 
\be 
B_z(t) = \frac{-\sigma_T t}{e} \left(\frac{F_\phi}{r}+\frac{\partial {F_\phi}}{\partial r}-\frac{1}{r}\frac{\partial F_{r}}{\partial \phi}\right).
\label{eq_bz}
\ee 
Here, $t$ is the time during which an electron remains in the system or within the radiation field. The integration constants are set to zero, as no magnetic field at $t=0$ is assumed. 

We stress that equations \ref{eq_br} -- \ref{eq_bz} are valid for short time scales. When the generated magnetic field becomes dynamically significant, additional terms that appear in Equation \ref{eq_mag_general} must be considered in determining the magnetic field evolution. 

The magnitude of the generated field is given by 
\be 
B = \sqrt{B_r^2 + B_\phi^2 + B_z^2}.
\label{eq_B_mag}
\ee 
To calculate the generation and the time evolution of magnetic fields above the disc, knowledge of all the components of radiative fluxes as well as their respective coordinate derivatives is needed.


\section{Computation of radiation flux and its rotation above the disc plane}
\label{sec_disc}
\begin {figure}
\begin{center}
 \includegraphics[width=9cm, angle=0]{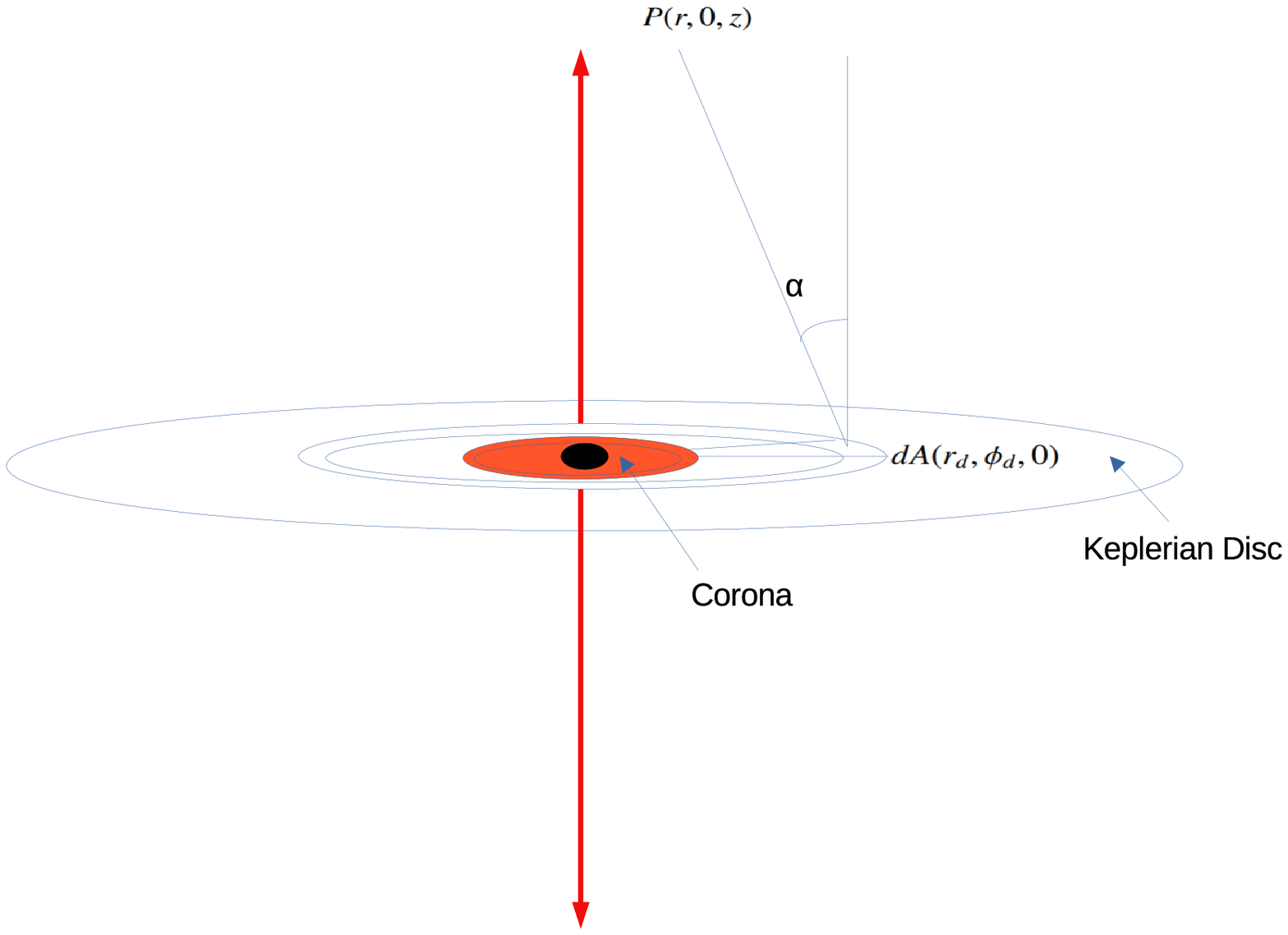}
\caption{Geometry of the accretion disc with two distinct components, an outer Keplerian disc and inner radiating corona separated at distance $r_d = r_s$. 
}
\label{lab_accretion_geom}
 \end{center}
\end{figure}
Consider a radiating, rotating accretion disc around a black hole. We denote here and below the radial coordinate of the disc location as $r_d$ (the radial distance from the black hole’s center), to discriminate it from the illuminating point, whose radial coordinate is denoted simply by $r$. 
The disc consists of two distinct radiating components: an outer Keplerian part at radii $r_d > r_s$, and an inner luminous corona at radii $r_d < r_s$. Here, $r_s$ represents the outer boundary of the corona, as is demonstrated in Figure~\ref{lab_accretion_geom}. We point out that the Keplerian disc is stable only for $r_d \gtrsim 3r_g$ ($r_g$ is the Schwarzschild radius), beyond which analytic solutions exist \citep[][]{1973A&A....24..337S}.  For simplicity, we approximate here the disc to be of negligible thickness (i.e., $z_d = 0$). 

The components of the radiative flux in the vicinity of the disc are given by
\be 
F^i = \int I l^i d\Omega.
\label{eq_flux}
\ee 
Here, $I$ denotes the specific intensity, and $l^i$ represents the corresponding direction cosine, from an illuminated point $P(r, 0, z)$ located above the disc (where the radiation moments are to be evaluated) to a surface disc element $dA(r_d, \phi_d, 0)$, with the corresponding differential solid angle $d\Omega$ (see Figure ~\ref{lab_accretion_geom}).

The coordinate derivatives of these fluxes are
\be 
\frac{dF^i }{dx_j} = \frac{d}{dx_j}\int I l^i d\Omega,
\label{eq_grad_flux}
\ee 
where $x_j \in \{ r,\phi,z \}$.
Note that the system possesses azimuthal symmetry, hence the derivatives of all components of fluxes with respect to $\phi$ are zero in Equations \ref{eq_br}-\ref{eq_bz}.
Denoting by $s$ the distance from the source area element to the illuminated point $P$, and using the assumption of infinitesimally thin disc, the differential solid angle is
\be 
d\Omega = \frac{dA \cos \alpha}{s^2} = \frac{zr_dd\phi_ddr_d}{s^3},
\ee 
with $\alpha$ being the angle between the field point and the disc normal, and $s =  \sqrt{r^2+r_d^2+z^2-2rr_d\cos \phi_d}$.
The specific intensity $I$ at point $P$ is obtained by Lorentz transforming the specific intensity $I_0$ of the Keplerian disc when matter is at rest \citep{1986rpa..book.....R, 2005MNRAS.356..145C, 2008bhad.book.....K},
\be 
I=\frac{I_0}{\gamma_d^4(1-u_il^i)^4}.
\ee 
Here, $\gamma_d$ is the local Lorentz factor of the rotating surface (relative to an observer at rest), and $u_i$ are the velocity components of the accretion disc normalized to the light speed, $c$. 

For disc around a non rotating black hole, in the pseudo-Newtonian regime, the dimensionless Keplerian velocity in units of $c$ is given by \citep{1980A&A....88...23P, 2008bhad.book.....K},
\be 
v_\phi = v_k = \sqrt{\frac{r_d'}{2(r_d'-1)^2}} {\rm~~for~~} r_d>r_s,
\ee
and for the inner corona, the azimuthal speed is  
\be 
v_\phi = v_c = \sqrt{\frac{r_s'}{2(r_s'-1)^2}}\frac{r_d}{r_s} {\rm~~for~~} r_d<r_s.
\label{eq:v_phi_corona}
\ee 
Here, $r_d'=r_d/r_g$ and $r_s'=r_s/r_g$ are the radii normalized to the Schwarzschild radius. 
In deriving equation \ref{eq:v_phi_corona}, we assume a constant angular velocity of the inner corona, similar to the results derived by \cite{2007MNRAS.375..513M}, which is expected for a strongly coupled plasma near the horizon of a rapidly rotating black hole. This assumption leads to a linear increase of $v_\phi$ with radius. We note that this assumption is different from that used in many GR-MHD simulations that consider an initial constant specific angular momentum, resulting in $v_\phi \propto r_d^{-1}$ \citep[at least in early stages of accretion][]{2004PhDT.........3M}.

Using these results, 
the components of the accretion disc velocity $u_i$ are $u_i = \{ -v_\phi \sin{\phi_d}, v_\phi \cos{\phi_d}, 0 \}$, while the direction cosines from the source point on the disc to the illuminated point $P$ are
\bea 
l^r= \frac{(r-r_d \cos{\phi_d})}{s}; ~
l^{\phi} = \frac{-r_d\sin \phi_d}{s} ; ~ 
l^z=\frac{z}{s},
\eea 
and $\gamma_d^2 = 1/(1-v_\phi^2)$. 

The specific intensity of the Keplerian disc is given by \citep{1973A&A....24..337S},
\be 
I_0 = I_k = \frac{3GM_B\dot{M}}{8\pi r_d^3}\left(1-\sqrt {\frac{3r_g}{r_d}}\right).
\label{eq:22}
\ee 
Here, $G$ is the gravitational constant and $\dot{M}$ is the mass accretion rate, given by
\(
\dot{M} = m \dot{M}_{\text{Edd}} = 1.4 \times 10^{17} m ~({M_{\rm BH}}/{M_{\odot}})~{\rm gr/s},
\)
where $\dot{M}_{\text{Edd}}$ is the Eddington accretion rate,
$M_{\rm BH}$ is the black hole mass and $M_{\odot}$ is the solar mass. 
Note that we assume here that the disc is truncated at an inner radius $r_s \geq 3r_g$. 


In our model, we assume that the radiation that escapes from the inner corona is approximately homogeneous. This is motivated by global radiation MHD simulations \citep{2019ApJ...885..144J}, which show that in the inner region, the corona becomes compact, hot, and characterized by a relatively uniform radiation energy density.
Hence, the specific intensity of the inner disc corona is assumed to be
\be 
I_0 = I_c = \frac{L_c}{A_c}.
\ee 
Here $L_c ( =l_c L_{Edd})$ is the corona luminosity and $L_{Edd} = \dot M_{\rm Edd} c^2 = 1.38 \times 10^{38}~{M_{\rm BH}}/{M_{\odot}}$~erg~s$^{-1}$ is the Eddington luminosity. The area of the corona (inner disc) is $A_c=\pi r_{s}^2$.
The disc (inner corona and outer Keplerian) radiates, thereby producing a radiation field above and below the disc plane.

\subsection{Available time for magnetic field growth}

Under the simplified assumptions explained in Section \ref{sec_ana}, the magnetic field initially grows linearly with time. The fields can grow only as long as the particles remain in the strong radiation field. An order of magnitude estimate of the available time can be made as follows.

The plasma above and below the accretion disc can either escape in the form of winds, it may fall into the disc, or it can spiral into the black hole. 
These processes operate on different characteristic timescales, typically of the order $\sim r_g / v$, where $v$ is the velocity of the matter. The material located above the corona can be expelled in the form of jets and winds. Right above the discs, the radiation-driven outflows exhibit velocities in the range of $10^{-4}c$ to $0.1c$, accelerating to relativistic speeds at larger distances \citep{2019MNRAS.482.4203V,2021MNRAS.501.4850R}. These speeds correspond to typical timescales between $\sim 10^{-3}$ seconds and $\sim 1$ second, for $M_{BH} \simeq 10 M_\odot$ \footnote{As we show below, when material in the wind changes its location on a typical scale of a few $r_g$, the magnetic field generation is strongly suppressed.}. Additionally, the infalling matter within the accretion disc evolves on a distinct timescale governed by viscosity, commonly referred to as the viscous timescale, which depends on the viscosity parameter $\alpha$. The typical viscous time during which matter at radial distance $R$ (using now spherical coordinates) falls into the black hole is $ t=t_v = R/{v_r} $. Here $v_r$ is the radial component of the plasma velocity which is \citep{1977A&A....59..111B},
\be 
v_r = 7.7 \times 10^{9}~ \frac{\alpha m^2 (1-r_d'^{-1/2})}{r_d'^{5/2}}~{\rm cm~s^{-1}}
\label{eq_vr}
\ee 
This velocity is about one order of magnitude smaller compared to \cite{1973A&A....24..337S} due to a larger density above the accretion plane, as shown by \cite{1977A&A....59..111B}.

 

\begin {figure}
\begin{center}
 \includegraphics[width=9cm, angle=0]{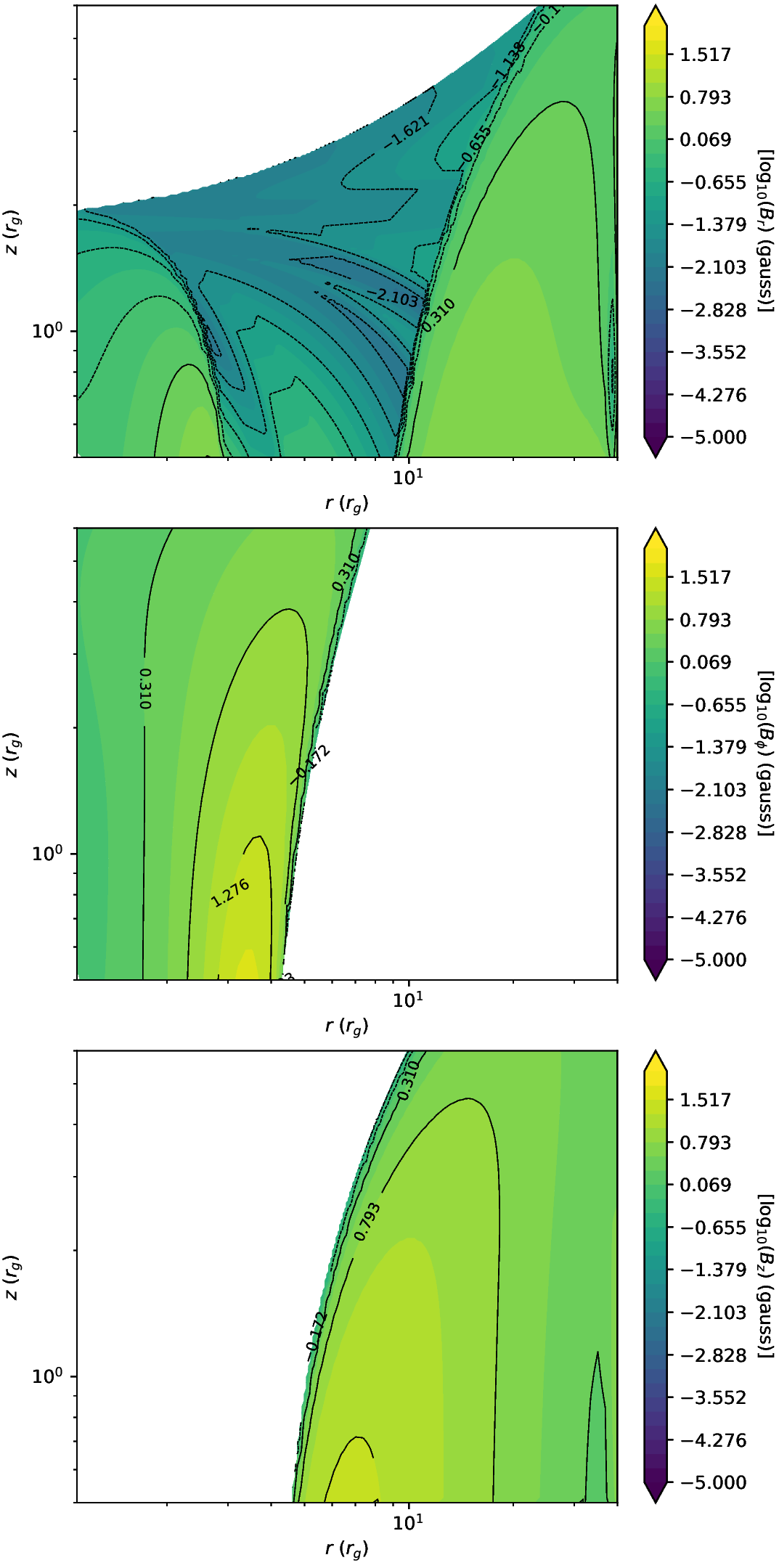}
 \caption{The three components of the magnetic field contours in $r-z$ plane for an accretion disc truncating at 3 Schwarzschild radius ($r_g$) when a corona component is absent. The mass of the black hole is $10~ M_\odot$ and the accretion rate considered is 1 Eddington accretion rate.}
\label{lab_contour_Br_mk_1_lc_0}
 \end{center}
\end{figure}

\begin {figure}
\begin{center}

 \includegraphics[width=9cm, angle=0]{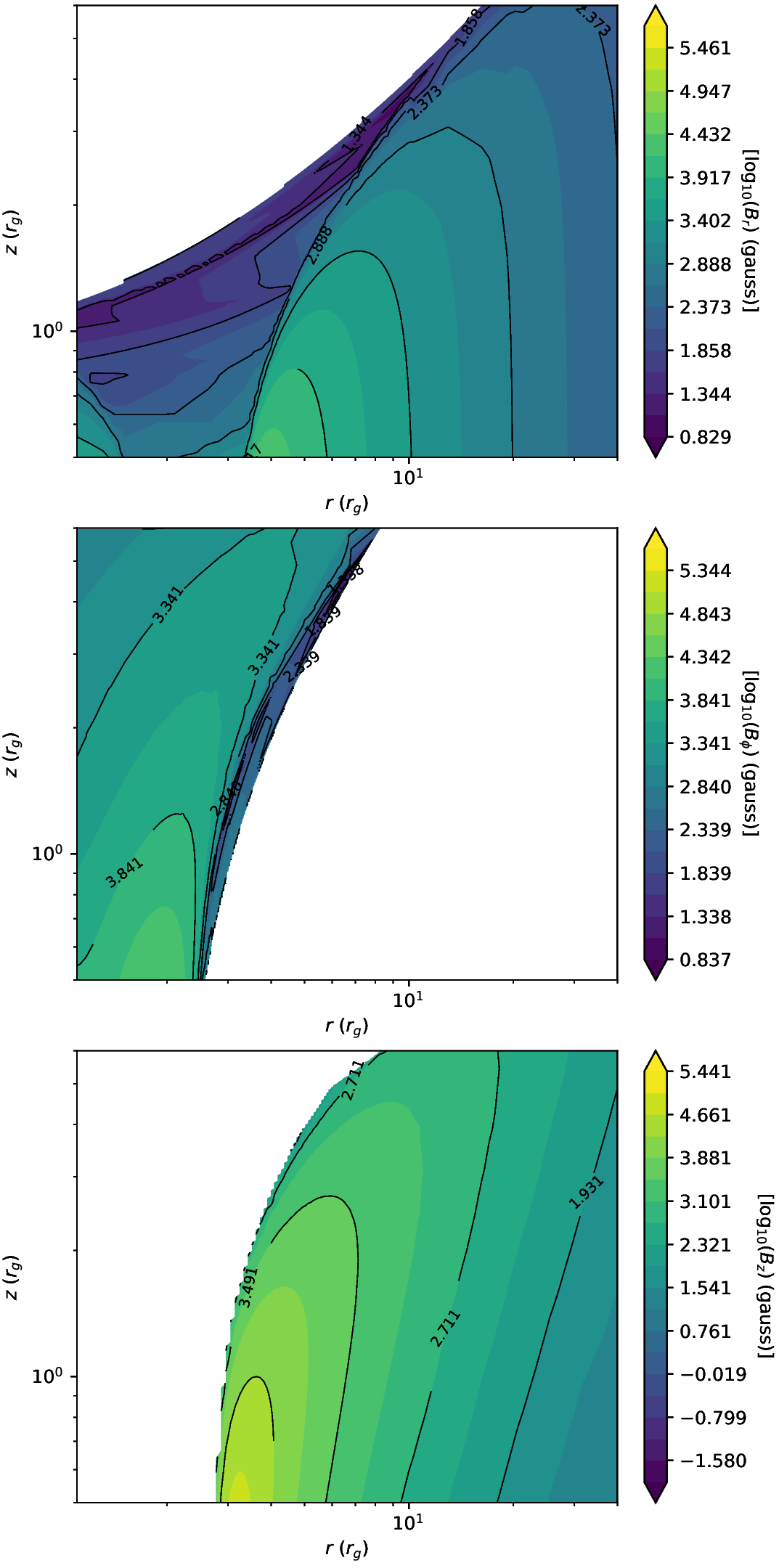}

\caption{Same as in Fig \ref{lab_contour_Br_mk_1_lc_0} but with the addition of a corona having luminosity $l_c=1$. Note the orders of magnitude difference in the magnitude of the magnetic field components relative to the results in Figure \ref{lab_contour_Br_mk_1_lc_0} }
\label{lab_contour_Br_mk_1_lc_1}
 \end{center}
\end{figure}

\begin {figure}
\begin{center}
  \includegraphics[trim={0cm 0cm 0.0cm 0cm},clip, width=8cm, angle=0]{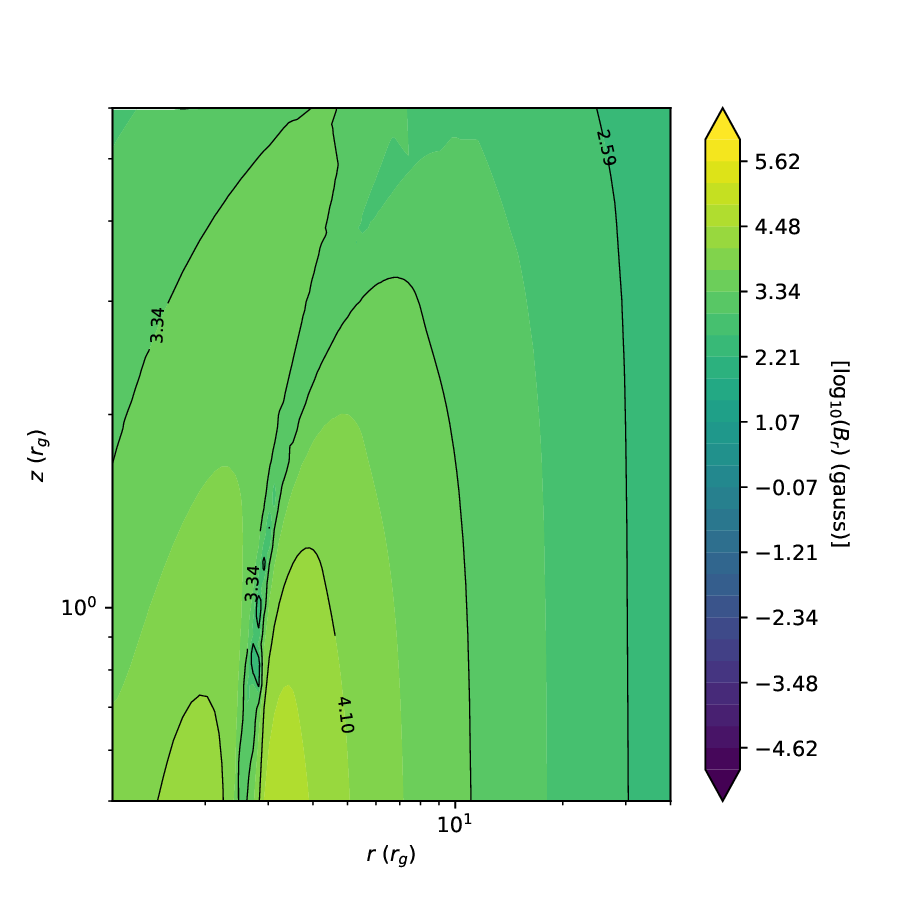}

\caption{Structure of the resultant magnetic field in the vicinity of the accretion disc corresponding to parameters in Figure \ref{lab_contour_Br_mk_1_lc_1}.
}
\label{lab_mag_field}
 \end{center}
\end{figure}

\begin {figure}
\begin{center}
 \includegraphics[trim={0cm 0cm 0.0cm 0cm},clip, width=8cm, angle=0]{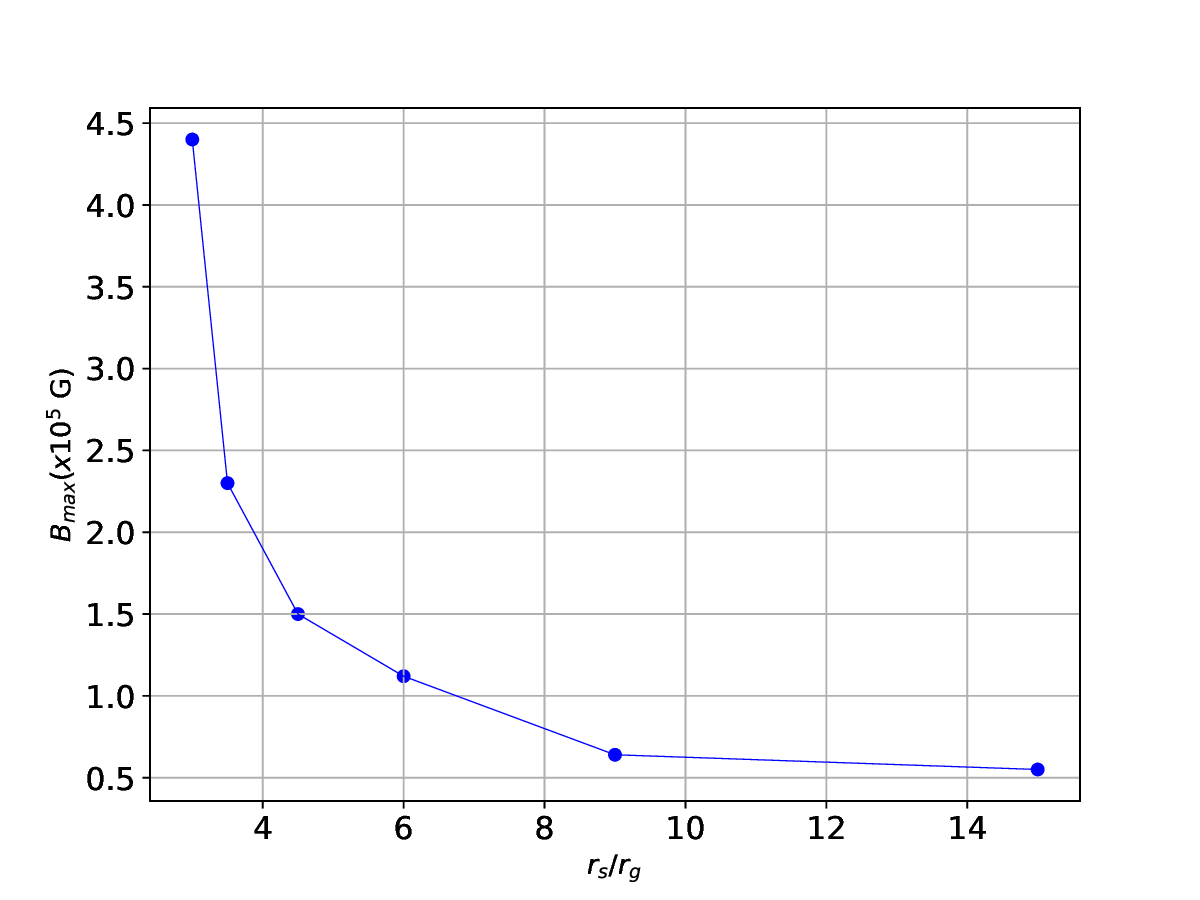}
  \includegraphics[trim={0cm 0cm 0.0cm 0cm},clip, width=8cm, angle=0]{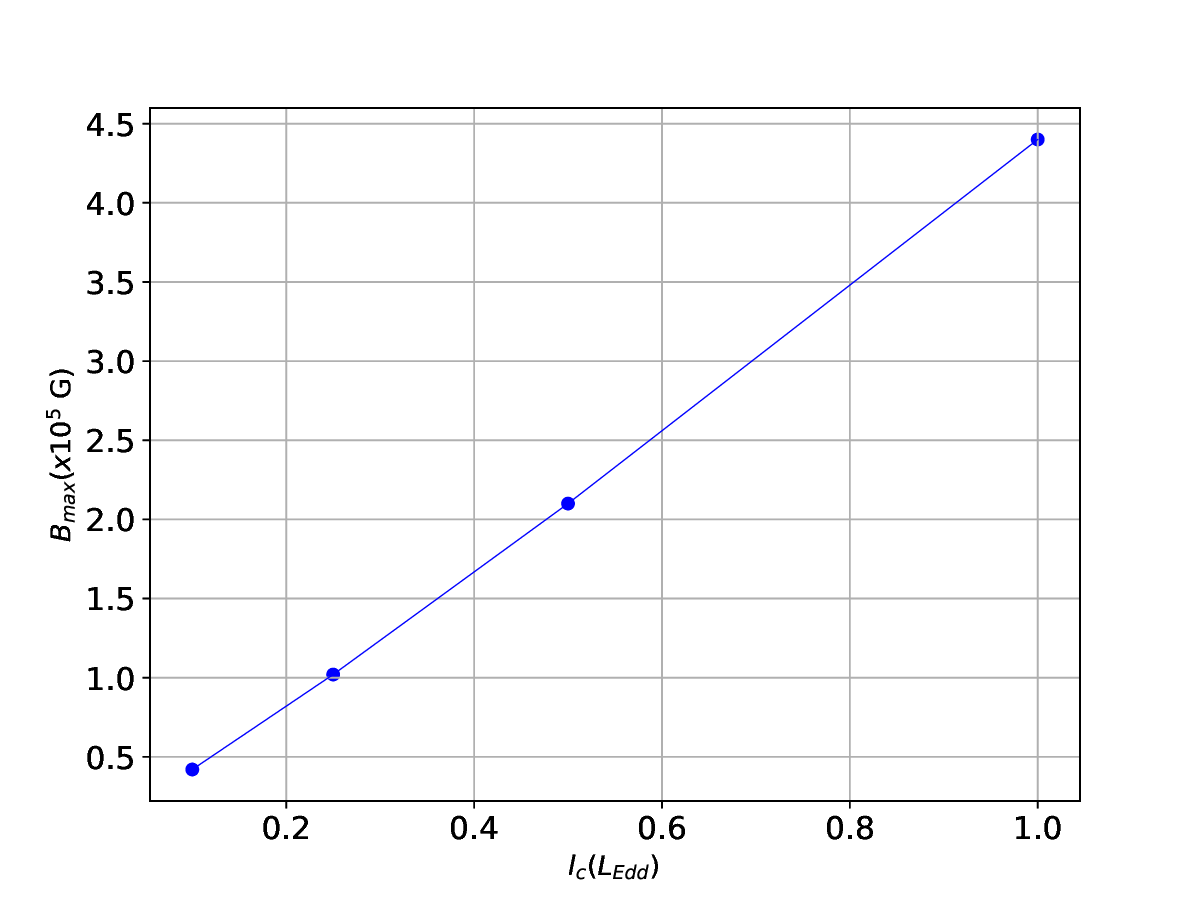}
\caption{Top: Dependence of the maximum magnetic field produced above the disc ($B_{\rm max}$) on the corona size ($r_s$), with other parameters the same as considered in section \ref{sec_res_cor_lum}. Bottom: dependence of the maximum field strength (Gauss) with corona luminosity $l_c$ given in Eddington units, for fixed corona size, $r_s = 3 r_g$.
}
\label{lab_B_rl}
 \end{center}
\end{figure}

Using the relations derived above, the three components of radiative flux are calculated by the set of equations \ref{eq_flux}, and then six components of the respective coordinate derivatives of these fluxes are calculated using Equation \ref{eq_grad_flux}. Putting them in equations \ref{eq_br}, \ref{eq_bfi} and \ref{eq_bz} we obtain the time evolution of the magnetic field components and collective three-dimensional magnetic field structure for the timescales $r_g/v_r$ from equation \ref{eq_vr}. \\ \\

\section{Results}
\label{sec_results}
We analyze the generation of magnetic fields above an accretion disc around a stellar mass black hole ($M_{BH}=10 {M_{\odot}}$) by solving the magnetic field induction equation under the influence of a non-conservative radiation field. The results are presented for two different configurations: (i) a standard Keplerian accretion disc without an inner corona, and (ii) a Keplerian disc with a luminous corona extending below $r_d=r_s$. The first of these serves as a reference for comparison with previous studies, while the second case with luminous corona investigates the impact of a strong radiation field from the inner corona.

\subsection{Magnetic Fields due to radiation from Keplerian Accretion disc}

In the absence of a luminous corona, we consider a Keplerian accretion disc with a dimensionless accretion rate $m = 0.1$ (corresponding to an Eddington-limited accretion rate) and an outer radius of $r_0 = 40 r_g$. This limit ensures an efficient calculation of radiative moments, as most radiation in the Keplerian disc comes from the inner region. In reality, the Keplerian disc extends to several orders of magnitudes larger than this value of $r_0$ we take, but this will not affect the results.
Figure \ref{lab_contour_Br_mk_1_lc_0} presents the spatial distribution of the $r$, $\phi$, and $z$ components of the generated magnetic field above the accretion disc. 
The produced magnetic field strength in this case remains relatively weak, reaching only a few Gauss, consistent with previous calculations by \citet{1977A&A....59..111B} and \citet{2002ApJ...580..380B}. These magnetic fields are several orders of magnitude lower than the equipartition values ($\sim 10^7$ Gauss), at which the magnetic energy density becomes comparable to the local gas pressure around stellar mass black holes \citep{1998ApJ...508..859C}.
Hence, they are not dynamically significant unless amplified by additional mechanisms, such as MRI turbulence or dynamo action. 

\subsection{Magnetic Fields with a Luminous Inner Corona}
\label{sec_res_cor_lum}

The presence of a luminous rotating corona significantly enhances the rate of magnetic field generation. We consider a scenario where the inner corona extends at $r_d < r_s = 3 r_g$ and radiates with a luminosity $l_c = 1$ (i.e., Eddington luminosity) while the outer Keplerian disc accretes at a sub-Eddington rate of $m = 0.1$. The spatial distributions of the resulting magnetic field components are shown in Figure \ref{lab_contour_Br_mk_1_lc_1}.
In this scenario, the generated magnetic field strength is significantly amplified, reaching values of up to $10^5$ Gauss near the inner disc surface, as shown in Figure \ref{lab_contour_Br_mk_1_lc_1}. 
This 
indicates that the presence of an intense radiation field introduces additional asymmetries that drive stronger field generation. As the growth timescale taken here is the viscous timescale $t(r) = r_g/v_r(r)$, the outcome contrasts with earlier findings that required unrealistically long growth timescales for significant magnetic field amplification \citep{1998ApJ...508..859C}.
This result demonstrates that the non-zero curl of the radiation force in the presence of a luminous corona provides an efficient mechanism for generating dynamically significant magnetic fields around black holes. 
The magnetic field strengths obtained here reach $\sim 5\%$ of the equipartition values typically required near black holes. Nonetheless, these values are significantly higher than those generated by a Keplerian disc alone.

In Figure \ref{lab_B_rl}, we show that the magnitude of the produced magnetic fields above the corona (Equation \ref{eq_B_mag}) decreases monotonously with the corona size, as $B_{\rm max} \propto 1/r_s^2$. Larger corona surface (assuming a constant luminosity) results in a smaller radiative flux and thus the produced magnetic field is reduced. 
In the bottom panel of Figure \ref{lab_B_rl}, we choose $r_s=3r_g$ and show that the maximum value of the produced magnetic field strength increases linearly with the corona luminosity, as the field is linearly proportional to the radiation fluxes emitted by the disc.

\section{Conclusion}
\label{sec_conclusions}
In this study, we have explored the generation of magnetic fields in the vicinity of black hole accretion discs due to the non-conservative nature of radiation forces. 
The presence of a structured, non zero curl radiative force naturally induces charge separation and subsequent magnetic field generation in the surrounding plasma. We have shown that this can lead to the formation of a dynamically important magnetic field, generated during feasible timescales in luminous radiative sources, such as inner disc corona.

Our findings indicate that the generated magnetic field in a standard Keplerian accretion disc remains at the level of a few Gauss, consistent with previous calculations \citep{1977A&A....59..111B, 2002ApJ...580..380B}. However, the presence of a luminous rotating corona, particularly at small radii ($r_d \sim 3 r_g$), significantly enhances the field growth, reaching magnitudes above $10^5$ Gauss. We affirm that the free parameters such as corona luminosity, the viscosity as well as the size of the corona do have quantitative effects on the produced fields. However, within the commonly assumed range of these parameters, the produced fields are strong enough to be dynamically significant. 

The magnitude of the generated magnetic fields is below equipartition value, typically at a level of a few percents of it.
The timescale for field growth is governed by the plasma inflow velocity. We show that for plausible values, this mechanism is viable within the typical lifetime of matter around accretion discs.  These results underscore the significance of this mechanism as a robust contributor to magnetic field generation in astrophysical plasmas. 

An important outcome of our analysis is the emergence of a dominant vertical component of the magnetic field, $B_z$, in the region above the inner corona (Figure \ref{lab_contour_Br_mk_1_lc_1}). This vertical dominance is significant for several reasons. First, a strong $B_z$ field is a key prerequisite for launching magnetically driven outflows and relativistic jets via the Blandford–Payne or Blandford–Znajek mechanisms. The presence of $B_z$ threading the disc-corona interface facilitates efficient extraction of angular momentum and vertical transport of energy. Second, it marks a clear deviation from purely toroidal field–dominated configurations, such as those typically generated by shear amplification alone in Keplerian discs. The appearance of a poloidally oriented field here reflects the influence of the non-zero curl of the radiation force, which naturally generates vertical current structures and hence $B_z$ dominance.

The presence of structured large scale strong magnetic fields near the accretion disc has observational implications.
The fields peak near the inner corona and decay faster than those typically produced by MRI or seed-field amplification. This may manifest as breaks in observed spectra (e.g., synchrotron or inverse-Compton emission) or radial variations in polarization. The dominance of vertical magnetic fields ($B_z$) can lead to high degrees of X-ray polarization with specific angular alignment, distinguishable via missions like Imaging X-ray Polarimetry Explorer (IXPE) and enhanced X-ray Timing and Polarimetry Mission (eXTP) \citep{2016SPIE.9905E..17W, 2019SCPMA..6229502Z}. These signatures offer testable avenues to distinguish this radiation-driven mechanism from conventional magnetic field origins.



While our analytical framework provides strong evidence for this process, further work is needed to explore its full impact under more complex conditions. As the estimations are based on various assumptions highlighted in section \ref{sec_ana}, future studies such as simulations of fields equations under general conditions (Equation \ref{eq_field_eq}) and numerical General Relativistic Magnetohydrodynamic (GRMHD) simulations incorporating radiation transport and non-ideal plasma effects will be essential to validate the theoretical predictions presented here. In future work, we aim to relax these assumptions and conduct a more general and comprehensive study.

In summary, we reinvestigate a novel and physically motivated mechanism for dynamically significant magnetic field generation around black hole accretion discs. The findings significantly enhance our understanding of accretion disc magnetization and open a new direction for studying the interaction between radiation and magnetic fields in astrophysical environments with extreme radiation fields.

\section*{Acknowledgments}
We acknowledge the support from the European Union (EU) via ERC consolidator grant 773062 (O.M.J.) and to the Israel Space Agency via grant \#6766. 

\bibliography{example}{}
\bibliographystyle{aasjournal}
\end{document}